
\input phyzzx

\Pubnum{ \vbox{ \hbox{Imperial TH/91-92/28} \hbox{QMW 92-9}
                    \hbox{hep-th/9207066} } }
\pubtype{}
\date{June 1992}

\titlepage

\title{The Global Phase Space Structure of the Wess-Zumino-Witten Model}

\author{G. Papadopoulos}
\address{Department of Physics \break Queen Mary and Westfield College\break
         London E1 4NS}
\andauthor{B. Spence}
\address{Blackett Laboratory\break Imperial College\break London
          SW7 2BZ\foot{Address from Oct. 1st 1992: Dept. Physics,
                      University of Melbourne, Victoria 3052 Australia}}

\abstract{We present a new parameterisation of the space of solutions of
the Wess-Zumino-Witten model on a cylinder, with target space a compact,
connected Lie group $G$. Using the covariant canonical approach
the phase space of the theory is shown
to be the co-tangent bundle of the loop group
of the Lie group $G$, in agreement with the result from the
Hamiltonian approach. The Poisson brackets in this phase space
are derived. Other formulations in the literature are shown to be obtained
by locally-valid gauge-fixings in this phase space.}

\endpage


\def\gm{{g^{-1}}}

\def\ua{{u^{-1}\delta u}}
\def\Va{{\delta V V^{-1}}}
\def\Ua{{U^{-1}\delta U}}
\def\va{{\delta v v^{-1}}}
\def\half{{1\over2}}

\def\cA{{\cal W}}
\def\pb#1#2{ \{#1,#2\} }
\def\pl{Phys. Lett.\ }
\def\cmp{Commun. Math. Phys.\ }
\def\np{Nucl. Phys.\ }
\def\uup{ \hbox{ \lower3ex\hbox{$\textstyle\otimes$} }}
\def\pbtimes#1#2{ \{{#1}\vbox{\uup\hbox{$\;\;,$}}{#2}\} }
\def\tr {{\rm tr}}

\def\lie{{\rm Lie}}
\def\LieG{{\rm Lie}G}
\def\CA {{\cal A}}
\def\kaffa{{\kappa\over4\pi}}
\def\kaape{{\kappa\over8\pi}}

\REF\cc{R. Abraham and J.E. Marsden,  {\it Foundations of Mechanics},
       Benjamin / Cummings  Publishing  Company, 1978;\hfill\break
	G. Zuckerman, in
         {\it Mathematical Aspects of String Theory}, ed. S.-T. Yau,
            World Scientific, Singapore 1987;\hfill\break
            \v C. Crnkovi\'c and E. Witten, in {\it Three Hundred Years of
           Gravitation}, eds. S.W. Hawking and W. Israel, C.U.P.
           Cambridge, 1987.}
\REF\witten{E. Witten, \cmp 92 (1984) 455.}
\REF\blok{B. Blok, \pl 233B (1989) 359.}
\REF\ash{A.Yu. Alekseev and S. Shatashvili, \cmp 128 (1990) 197; 133
        (1990) 353.}
\REF\faddeev{L.D. Faddeev, \cmp 132 (1990) 131.}
\REF\balog{J. Balog, L. D\c abrowski and L. Feh\'er, \pl 244B (1990) 227.}
\REF\felder{G. Felder, K. Gaw\c edski and A. Kupiainen, \np B299 (1988)
        355; \cmp 117 (1988) 127.}
\REF\gawedski{K. Gaw\c edski, \cmp 139 (1991) 201.}
\REF\bimonte{G. Bimonte, P. Salomonson, A. Simoni and A. Stern, {\it
             Poisson Bracket Algebra for Chiral Group Elements in
              WZNW Model}, preprint UAHEP 9114.}
\REF\chu{M.F. Chu, P. Goddard, I. Halliday, D. Olive and A. Schwimmer,
                \pl B266 (1991) 71.}
\REF\us{G. Papadopoulos and B. Spence, {\it The Canonical Structure
            of Wess-Zumino-Witten Models}, preprint Imperial
             91-92/19, QMW 92/2.}
\REF\faddeevtwo{A.Yu. Alekseev and L.D. Faddeev, {\it $(T^*(G))_t:$ A
            Toy Model for Conformal Field Theory}, Commun. Math.
             Phys. to appear;\hfill\break
             L.D. Faddeev, {\it Quantum Symmetry in Conformal
              Field Theory by Hamiltonian Methods}, Cargese Lectures 1991.}
\REF\hwzw{I. Bakas and D. McMullan, \pl B189 (1987) 141.}
\REF\gribov{V.N. Gribov, \np B139 (1978) 1;
             \hfill\break I. Singer, \cmp 60 (1978) 7.}

\sequentialequations


\chapter{Introduction}

It has been known for a number of years that the phase space of a classical
system can be defined in two different ways.  The first is the
Hamiltonian definition of the phase space as the space of positions and
momenta of the system; we will call this phase space $P_H$.  In the second
definition, the phase space of a classical system is defined as the space
of solutions of its Lagrangian equations of motion (see Ref. [\cc] for
discussions).  This  `covariant canonical' definition of the phase space
gives a space we will call $P_C$. Although it is expected that these two
definitions for the phase space of a classical system give spaces that are
locally equivalent, globally they may be quite different spaces. However,
as we will see
below in the case of the Wess-Zumino-Witten (WZW) model, these two phase spaces
are the same.

The covariant definition of the phase space is particularly suited to
theories for which the general solution of the classical equations of motion is
known, and it has been applied extensively to two-dimensional conformal
field theories. Much work has been done on the canonical structure of
the (WZW) models recently [\witten-\faddeevtwo], and in particular
on the covariant approach to the phase space. This has included study of the
symplectic structure
of the WZW model, defined on the cylinder and with target space a
compact, connected Lie group $G$.  To construct the phase space $P_C$ of
this WZW model, it is necessary to introduce a parameterisation of the
space of solutions of the model.  Then the phase space $P_C$ is defined as
the space of independent parameters necessary to describe these classical
solutions.
In Ref. [\witten] the phase space of the theory was taken
to be $LG\times LG$, where $LG$ is the loop group of the group $G$.
In Ref. [\chu] the covariant approach led to the phase space
$P_C$ being diffeomorphic to ${LG\times LG\times G \over G}$.  An important
feature of this parameterisation was the presence of a symmetry in the
space of solutions, generated
by $G$.  However, neither of these phase spaces are
 diffeomorphic to the phase space of the WZW model defined in the Hamiltonian
approach (see Refs. [\gawedski, \hwzw]).
The latter, as we shall see, is the co-tangent bundle of the loop
group $LG$, \ie\ $P_H=T^*LG$.

In this paper, we will introduce a
new parameterisation of the space of solutions
of the WZW model.  The associated phase space $P_C$ is diffeomorphic
to ${LG\times LG\times {\CA} \over LG}$ where $\CA$ is the space of
connections over the circle. In this parameterisation
the solution space
symmetry generated by the group $LG$ can be fixed completely;
additionally  the corresponding Poisson brackets of the WZW model are easily
derived. Furthermore, the resulting phase space
$P_C$ is diffeomorphic to the phase space $P_H$ given in
the Hamiltonian treatment of this model.  We will see how phase spaces studied
in the literature can be recovered from our phase space by partially gauge
fixing the loop group symmetry. However, such
gauge-fixings are local and cannot
be extended globally because of  Gribov-Singer ambiguities. This is why the
resulting phase spaces are not diffeomorphic to those obtained by a global
gauge-fixing, and the resulting theories are in general different.

This paper has been organised as follows:  In Section Two, we will discuss
the example of a particle moving upon a group manifold.
This proves to have many of the features of the WZW
model, in a simpler form.  Both covariant and Hamiltonian
phase spaces are discussed. The Poisson brackets
of the covariant phase space model are
calculated by two different methods. In the first method we perform a
gauge fixing of the symmetries of the space of solutions of the model
and then calculate the Poisson brackets of the theory.
In the second method we will enhance
the phase space of the model in such a way that the symplectic form becomes
invertible, and then we will impose a set of first class constraints.
In Section Three, we present the corresponding discussion for the WZW
model. We begin by presenting
a new parameterisation of the solutions of the
WZW model. The resulting phase space is shown to be the co-tangent bundle of
the loop group, and we calculate the corresponding Poisson brackets,
first by gauge
fixing, and second by enhancing the phase space of the theory and
imposing constraints, as in the
particle model.
Finally, in Section Four, we relate our formulation of the phase space
of the WZW model to those appearing previously in the
literature, and we discuss the global issues which are relevant
to this relationship.


\chapter{The Particle on a Group Manifold}

The case of the particle moving upon a group manifold has been
recently discussed by Alekseev and Faddeev [\faddeevtwo]. We would like
to rephrase and extend this discussion for our purposes here. It will
turn out that this discussion is a close analogue of the corresponding
WZW discussion.
The action of this particle model is
$$
 S = \half\!\int\!dt\,\tr\,\big(\dot g\gm\big)^2, \eqn\aone
$$
where $\dot{g} = {d\over dt}g $ and  $g$ is a map from the real line $\Re$
into $G$. ${\rm Lie}G$
is the Lie algebra of the compact,
connected Lie group $G$, with a basis $\{t_a\}$ satisfying
$[t_a,t_b] = {f_{ab}}^ct_c$ and $\tr(t_at_b)=\delta_{ab}$\
$(a,b,=1,2,\dots,{\rm dim}(\LieG))$.
We have introduced a matrix
representation for the group $G$ and the multiplications in Eqn. \aone\
are matrix multiplications.
The equations of motion of the action \aone\ are
$${d\over dt}(\dot g\gm) = 0.  \eqn\atwo $$
The Hamiltonian treatment of this model is straightforward.
For convenience we introduce a local
parameterisation $X^i, i=1,2\ldots,{\rm dim} G$, of the group $G$.
We have $\dot g\gm = R^a_i\dot X^it_a$, with $R^a_i$ the right frame
of $G$. Writing the action \aone\ in these coordinates gives
$S=\half\int\!dt\, g_{ij}\dot X^i\dot X^j$, where $g_{ij}=R^a_iR^b_j
\delta_{ab}$ is the bi-invariant metric on the group manifold. Performing the
usual Hamiltonian analysis, one finds that the phase space $P_H$ is the
co-tangent bundle $T^*G$, with co-ordinates $(X^i,P_j)$, where
$P_j$ is the momentum and $X^i$ is the position of the particle.
The symplectic form  $\omega$ is equal to $\delta X^i\delta P_i$ and the
canonical Hamiltonian is $H = \half g^{ij}P_iP_j$.  The Poisson brackets are
$\pb {X^i} {P_j}=\delta^i_j$ (note that this fixes our conventions for deriving
the  Poisson brackets from the symplectic form).

Now we consider the covariant approach to the definition of the phase
space of this particle model. The equations of
motion \atwo\ are solved by
$$
   g = u e^{ta} v,     \eqn\athree
$$
with $u,v$ time-independent group elements and $a$ a time-independent
element of
$(\lie G)^*$. The group action
$$
    u\rightarrow u h,\quad a \rightarrow h^{-1} a h, \quad
       v\rightarrow h^{-1} v, 				\eqn\afour
$$
with $h\in G$, leaves the solution \athree\ invariant.

The phase space $P_C$ of the theory is the space of elements
 $\{u,v,a\}$, modulo the group action of Eqn. \afour.
$P_C$ is thus the coset space ${G\times G\times(\lie G)^*\over G}$, which is
diffeomorphic to the co-tangent bundle $T^*G=G\times(\LieG)^*$ of the group
$G$, in agreement with the Hamiltonian approach.
The symplectic form of $P_C$ in this approach is
$\omega= \delta X^i\delta(\partial S/\partial\dot X^i)$, giving
$$
        \eqalign{\omega &= \tr\Big(\delta g\,\delta (\gm\dot
                           g\gm)\Big)\cr
                     &=\tr\Big( (\gm\delta g){d\over dt}(\gm\delta g)
                                +(\delta g\,\gm){d\over dt}(\delta
                           g\,\gm)\Big).   \cr}    \eqn\asix
$$
$\omega$ is closed, $\delta\omega=0$, and time independent, $\dot\omega=0$ (we
take $t=0$ in the following). Now we write $\omega$ in terms of the phase
space variables $\{u,v,a\}$, giving
$$
   \omega =\tr\Big((\ua)^2\, a + (\ua)\,\delta a - (\va)^2\, a + (\va)\,\delta
a\Big).
         \eqn\aseven
$$
The Hamiltonian in these coordinates is $H = \half\tr a^2$.
The action \aone\ is invariant under the action
$g\rightarrow lgr^{-1}$, $l,r \in G$, of the
group $G$. In terms of the phase space variables $\{u,v,a\}$ the corresponding
currents (charges) are $j_l = uau^{-1}$ and $j_r = -v^{-1}av$.  These
currents are the particle model analogues of
the left and right currents of the WZW model.

The symplectic form \aseven\ is degenerate along the directions of the group
action \afour, and hence is not invertible. To calculate the Poisson brackets
of the theory, we may adopt one of two approaches. The
first is to gauge fix the symmetry generated by the group action
\afour. In the second, we enhance the phase space and invert the
symplectic form on the enhanced phase space.  Then we impose a first class
constraint necessary for the system to have the right number of degrees
of freedom. The constraint generates gauge transformations, and one
considers
gauge-invariant functions on the reduced phase space.


In the gauge-fixing approach, one can simply fix the symmetry \afour\
by using it to set $u=e$, where $e$ is the identity element of the
group $G$  (alternatively one could set $v =e$, with
equivalent results).  The gauge choice $u=e$ is a ``good'' gauge choice
because $G$ acts freely and transitively on the space of $u$'s, $\{u\}$. There
are thus
no residual symmetries. The symplectic form \aseven\ in the gauge $u=e$
is
$$
  \omega = -\half {f_{ab}}^c R^a_iR^b_j\,a_c\,\delta X^i \delta X^j
      + R^b_i\,\delta X^i\,\delta a_b, 			\eqn\aeight
$$
where $X^i$ are the group manifold coordinates corresponding to $v$.
This symplectic form is the standard symplectic form on $T^*G$ expressed in
terms of the right trivialisation of $T^*G$ and is the same as the symplectic
form of the phase space $P_H$ in the Hamiltonian treatment of the
model, with $P_i= R_i^b\,a_b$.
Similarly the symplectic form of Eqn. \aseven\ in the gauge $v=e$ is the
standard symplectic form of $T^*G$, this time expressed in terms of the left
trivialisation of $T^*G$.   The form in Eqn. \aeight\ is trivial to
invert, giving the Poisson brackets
$$
   \pb{X^i}{X^j} = 0, \quad \pb{X^i}{a_b} =
                          R_b^i,\quad
      \pb{a_a}{a_b} = {f_{ab}}^c a_c.  			\eqn\anine
$$
{}From these brackets it is straightforward to deduce
the Poisson brackets of functions of $X$ and $a$.
For example, in this gauge the Poisson brackets
of the currents $j_l=a$, $j_r=-v^{-1}av$, give
two commuting copies of the Lie algebra $\lie G$, \ie\ the Poisson bracket
of each current with itself is the same as the Lie bracket
of $\lie G$ and $\{j_l, j_r\}=0$.


The other way to invert the symplectic form \afour\ is to extend the phase
space of the particle model, and then to impose constraints.
We define an extended model -- the `lr' model -- to have
a phase space $P_{lr}$
parameterised by $(u,v,a_l,a_r)$, with $u,v\in G$ and $a_l,a_r\in
(\lie G)^*$, \ie\ $P_{lr}=G\times G\times (\LieG)^*\times (\LieG)^*$.
The symplectic form is $\omega_{lr} = \omega_l + \omega_r$, where
$$
  \omega_l =\tr\Big((\ua)^2\,a_l + (\ua)\,\delta a_l\Big),\quad
  \omega_r = \tr\Big(-(\va)^2\,a_r + (\va)\,\delta a_r\Big). \eqn\afourteen
$$
These forms are closed and time independent. The Hamiltonian of the system
we take to be $H_{lr} = {1\over 4}\tr(a_l^2 + a_r^2)$.
The symplectic form $\omega_{lr}$ is non-degenerate,
and hence invertible. The forms $\omega_l$ and $\omega_r$ can be
inverted separately.  In addition the form $\omega_r$ is the same as the
form $\omega$ of Eqn.
\aeight\ and it can be inverted in the same way.  Similarly we can invert
the form $\omega_l$.  Since the symplectic form $\omega_{lr}$
factorises, all the
Poisson brackets of the
$\{u, a_l\}$ variables with the $\{v,a_r\}$ variables vanish.
To recover the particle model from the $lr$ model, we impose the constraint
$Q=a_r-a_l = 0$. This is a first class constraint, whose Poisson bracket
with $H_{lr}$ does
not generate
additional constraints. The subspace of $P_{lr}$ satisfying the
constraint $Q=0$ is isomorphic to $G\times G\times(\LieG)^*$, and the gauge
transformations generated by the constraint mod this by the group $G$,
and thus
the reduced phase space is isomorphic to $G\times(\LieG)^*$.
Thus the Hamiltonian
approach and the covariant canonical approach (with either
gauge-fixing or constraints) yield the same answer for the particle
model.


\chapter{The WZW Model}

There is a direct correspondence between the particle
model described in the previous
section, and the WZW model. This will lead us to a new
parameterisation of the space of solutions of the WZW model, with
a corresponding definition of
the phase space, and a derivation of the
Poisson brackets.

The equations of motion of the WZW model are
$$
       \partial_-(\partial_+g\,\gm) = 0.  \eqn\bone
$$
where $g$ is a map from a cylinder $S^1\times \Re$ to a compact,
connected Lie group $G$. The pairs
$(x,t): 0\leq x<1, -\infty<t<\infty$ are the
co-ordinates of $S^1\times \Re$ and we take $x^\pm = x \pm t, \partial_\pm =
\half(\partial_x \pm\partial_t)$ (note that these conventions differ
from those of Ref. [\us]).
The semilocal transformations $g\rightarrow
l(x^+) g r^{-1}(x^-)$, with $l,r$ maps from $S^1$ into $G$, are
symmetries of the theory;
the corresponding currents are
$$
J_+= -\kaffa\partial_+g g^{-1},  \qquad J_-= \kaffa g^{-1}
                          \partial_-g.    \eqn\bbone
$$
where $\kappa$ is the coupling constant of the WZW model.

In the Hamiltonian approach, one may consider the WZW model as a
two dimensional non-linear sigma model with Wess-Zumino term,
whose target space is the
manifold of the group $G$. Applying the usual Hamiltonian analysis to
this sigma model action, one finds directly that
the phase space of the WZW model is the co-tangent
bundle of its configuration space $LG$, \ie\ $P_H=T^*LG$.

Next we consider the covariant approach to the phase space.  The symplectic
form of the WZW model is (see Ref. [\chu], for example)
$$
    \Omega = -\kaape\,
   \int_0^1\!\!dx\,\tr\,\Big((\gm\delta g)\partial_+(\gm\delta g)
                 - (\delta g\,\gm)\partial_-(\delta g\,\gm)\Big).
             \eqn\btwo
$$
This symplectic form is closed and time independent (we take $t=0$ in
the following).
The key step we take at this juncture is to parameterise the space
of solutions to the field equations \bone\ in the following manner:
$$
   \eqalign{   g(x,t) &= U(x^+) \cA(A;x^+,x^-) V(x^-), \crr
                 \cA(A;x^+,x^-) &=
                     P\exp\!\int_{x^-}^{x^+}\!A(s)ds,\cr }
						\eqn\bthree
$$
where $U$ and $V$ are maps from $S^1$ to the group $G$, and
the field $A$ in the path-ordered exponential is a $\LieG$-valued
connection
\foot {to be precise, $A$ is a $(\LieG)^*$-valued periodic one-form on
the real line, but for simplicity we have identified
$\LieG$ with its dual using
the invariant metric on $\LieG$} over $S^1$. The fields $U$, $V$ and $A$
are thus periodic in $x$. The expression for $g(x,t)$ in Eqn. \bthree\
is then periodic in $x$ and solves
the field equations \bone.  The latter result
follows immediately from the fact that $\cA$
satisfies the parallel transport equation
$$
        \partial_s\cA (A;s,x^-) = A(s)\ \cA (A;s,x^-).
						\eqn\bfour
$$
This equation implies that
$\partial_+\cA = A(x^+) \cA$, $\partial_-\cA=-\cA A(x^-)$.
 Inserting the solution \bthree\ into the symplectic form \btwo\ gives
$$
   \eqalign{
    \Omega = -\kaape\,\int_0^1\!dx\,\tr\,\Big( &(\Ua)\partial_x(\Ua)
              + 2(\Ua)^2\,A + 2(\Ua)\delta A \cr &
      - (\Va)\partial_x(\Va) -2(\Va)^2\,A
+ 2(\Va)\delta A \Big).  \cr}       \eqn\bfive
$$
The solution $g$ of the WZW equations of motion given in the
parameterisation \bthree\ is invariant under the
transformations
$$\eqalign{
      U(x)&\rightarrow U(x) h(x),
         \quad V(x)\rightarrow h^{-1}(x) V(x),
\cr
A(x) &\rightarrow
-h^{-1}(x)\partial_x h(x)+ h^{-1}(x)A(x)\,h(x),}  \eqn\bsix
$$
where $h\in LG$.
To prove this, we observe that under these transformations
$\cA(A;x^+,x^-)\rightarrow h^{-1}(x^+) \cA(A;x^+,x^-) h(x^-)$.
The phase space $P_C$ of the WZW model is then the space of fields
$\{U,V,A\}$, modulo the  transformations \bsix.
This is ${LG\times LG\times
{\CA} \over LG}$ where $\CA$ is the space of $G$-connections over the
circle. This is diffeomorphic to $T^*LG$, \ie\ it is the same as the phase
space $P_H$ derived from the Hamiltonian treatment of the theory.

The symplectic form \bfive\ is degenerate along the directions of
the action \bsix\ of the loop group $LG$.
We may deal with this by gauge-fixing or by imposing
constraints.


We first consider the gauge-fixing approach and analogously to
the case of the particle
model we may choose as a gauge fixing condition  $U = e$ where $e$ is the
identity element of the loop group $LG$. This is a good gauge choice,
as $LG$ acts freely and transitively on the space of $U$'s, $\{U\}$. The
symplectic form \bfive\ then becomes
$$
    \Omega = -\kaape\,\int_0^1\!dx\, \tr\,\Big( -(\Va)\partial_x(\Va)
            - 2(\Va)^2\,A + 2(\Va)\delta A
         \Big).   \eqn\beight
$$
This symplectic form is not degenerate and is invertible. The
simplest way to invert it is to first rewrite it in terms of a local
parameterisation $X^i(x)$ for the maps $V$ ($V=V(X)$). This gives
$$
   \Omega = -\kaape\int_0^1\!dx\Big( -(R^a_i\delta X^i)\partial_x(R^a_j\delta
             X^j) - {f_{ab}}^c R^a_iR^b_jA_c\,\delta X^i\ \delta X^j
      + 2 R^a_i\, \delta X^i\delta A_a\Big).   \eqn\bnine
$$
The remarkable feature of this expression for the
form $\Omega$ is that one does not
need to invert any differential operator in order to invert the form
({\it c.f.} Refs. [\chu,\us], where in order to invert the symplectic
form it was necessary to find the inverse of
the operator $\partial_x$ on the circle).
Like the case of the particle model, the gauge
$U=e$ parameterises the symplectic form on $T^*LG$ in terms of the right
trivialisation of $T^*LG$ and the gauge $V=e$ parameterises the same
symplectic form in terms of the left trivialisation. The
inversion of the form \bnine\ is straightforwardly carried out, and
leads to the Poisson brackets ($\beta = -{4\pi\over\kappa}$) $$
     \eqalign{ \pb{X^i(x)}{X^j(y)} &= 0,\crr
                   \pb{X^i(x)}{A_a(y)}
                  & =\beta R^i_a[X(x)] \delta(x,y), \crr
       \pb{A_a(x)}{A_b(y)} &= \beta\Big(\delta_{ab}\partial_x +
                     {f_{ab}}^c A_c(x) \Big)\delta(x,y), \cr}
              \eqn\bten
$$
where $\delta(x,y)$ is the delta function on $S^1$.
The brackets \bten\ are the
Poisson brackets on the co-tangent bundle of the loop group
which one would expect --
here we have {\it derived} them from the WZW model, using the
corresponding symplectic form in the
parameterisation \bthree.
Using Eqn. \bten, we can calculate Poisson brackets involving $V$ and
$A$ -- for example $\pbtimes{V(x)}{V(y)} = 0$, $\pb{V(x)}{A_a(y)} =\beta
V(x)t_a\delta(x,y)$. In this gauge, the WZW currents \bbone\ become
$J_+ = -{\kappa \over 4\pi} A$, $J_- = {\kappa \over 4\pi}
(V^{-1}\partial_xV - V^{-1}AV) $, and it can be verified by a
straightforward calculation that their Poisson bracket algebra is isomorphic
to two commuting copies of a Kac-Moody algebra with a central extension.

In the constraint approach to the degeneracy of the form \bfive, we
introduce an `LR' model, with phase space
$P_{LR}= LG\times LG \times \CA \times \CA$, with coordinates
$(U,V,A^L,A^R)$. The symplectic form on $P_{LR}$ is defined
to be $\Omega_{LR}=\Omega_L
+ \Omega_R$, where
$$
 \eqalign{ \Omega_L &= -\kaape\!\int_0^1\!dx\,\tr \Big(
(\Ua)\partial_x(\Ua) + 2(\Ua)^2\,A^L +
                      2(\Ua)\delta A^L
         \Big),\cr
            \Omega_R &= -\kaape\!\int_0^1\!dx\,\tr\Big(-(\Va)\partial_x(\Va)
                  -2(\Va)^2\,A^R +
         2(\Va)\ \delta A^R
         \Big).\cr}    \eqn\beleven
$$
The symplectic form \beleven\
is similar to the symplectic form of Eqn. \bfive\ but
with different connections in
the $U$ and $V$ sectors.  It is straightforward to invert the forms
in Eqn. \beleven.
Indeed, the form $\Omega_R$ is the same as the symplectic form of Eqn. \beight\
which we have already inverted in the gauge-fixing method. The symplectic form
$\Omega_L$ can be treated in a similar way.  Poisson brackets
of variables of the
$U$ sector with variables of the $V$ sector are zero, because the
symplectic form $\Omega_{LR}$ factorises. For completeness we give the
Poisson brackets of
the $U$ sector:
$$
 \eqalign{
\pb{Y^i(x)}{Y^j(y)} &= 0,\quad \pb{Y^i(x)}{A^L_a(y)}
                   =\beta L^{i}_a[Y(x)]\delta (x,y),
\crr
       \pb{A^L_a(x)}{A^L_b(y)} &= \beta\Big(-\delta_{ab}\partial_x -
                     {f_{ab}}^c A^L_c(x) \Big)\delta (x,y), \cr}
              \eqn\btwelve $$
where $U=U(Y)$, with $Y$ a parameterisation of $LG$ in terms of local
co-ordinates on the group manifold $G$.
The Hamiltonian of the $LR$ model is taken to be
$H = {1\over4}\tr\int_0^1\!dx(J_L^2+J_R^2)$, where $J_L =-{\kappa \over 4 \pi}
(\partial_xU\,U^{-1} + UA^LU^{-1})$ and $J_R = {\kappa \over 4 \pi}
(V^{-1}\partial_xV - V^{-1}A^RV)$. To recover the WZW model we introduce the
constraint  $Q=A^R-A^L$ in the
phase space $P_{LR}$ of the LR model. Using the above Poisson
brackets, it follows that this constraint is first class
and its Poisson bracket with $H_{LR}$ does not induce any other
constraints. The Poisson bracket algebra of
the constraints is a Kac-Moody algebra without a central
extension.  The reduced phase space that we get by factoring out the
transformations on $P_{LR}$ generated by the constraints $Q$ is the same
as $T^*LG$, thus giving agreement with both the gauge-fixed
covariant approach and the Hamiltonian approach.


\chapter {Discussion}

We would now like to discuss how our results relate to other
work in the literature.
The symmetries of the space of solutions of the particle model (Eqn. \afour)
and
the  WZW model (Eqn. \bsix) can be treated by choosing
the gauge-fixing conditions to be
different from those considered in Sections Two and
Three above.  Using these other gauge-fixing conditions, we
can make contact with the
parameterisations of the spaces of solutions of these models in
Refs. [\chu,\faddeevtwo]. However, these gauge-fixings suffer from
Gribov-Singer-type ambiguities.  Because of
this, the resulting spaces of parameters are topologically different
from those obtained from the Hamiltonian treatments of these models.  In the
following for simplicity we assume that the group $G$ is simply connected.

For the particle model, the  symmetry Eqn. \afour\ can be
partially gauge-fixed by putting $a$ in the Cartan subalgebra
${\bf h}\subset\LieG$.
There is a residual symmetry associated with this gauge fixing.
This is $u\rightarrow u T$, $v\rightarrow T^{-1} v$, $a \rightarrow a$
($a\in \bf {h}$) where $T$ is in a maximal torus $\bf{T}$ of the group
$G$.  This parameterisation of the particle model was studied in Ref.
[\faddeevtwo].
Apart from the residual symmetry which must still be fixed, there is a
Gribov-Singer ambiguity associated with this gauge fixing.  One way to
see this is to observe that the space of independent parameters that
describes the solutions of the particle model, after introducing the above
gauge-fixing, is ${G\times G\times \bf{h} \over \bf{T}}$. This space is
{\it not} diffeomorphic to the phase space $T^*G$ of the
particle model -- for example, $\pi_2(T^*G) = 0 \not= \pi_2({G\times
G\times{\bf h}\over{\bf T}})$. This results from the fact that  this
gauge-fixing condition is local, and cannot be extended globally.  This is in
contrast to the gauge-fixing which we used in Section Two.

Similar comments apply to the WZW model. In our parameterisation \bthree\ one
can gauge-fix the connection $A$ so that it is a {\it constant}
connection over the circle.  The residual
transformations for this gauge-fixing are the constant gauge transformations.
The constant gauge transformations are parameterised by the elements of
the group $G$ and they act on the parameters of the solutions as
$U\rightarrow U k$, $V\rightarrow k^{-1} V $ and $A\rightarrow k^{-1}A k$
where $k\in G$ and $A$ is a constant connection.  This parameterisation of the
space of solutions is that of Ref. [\chu], and
the resulting phase space of the theory is ${LG\times LG
\times\LieG\over G}$.  This phase space is not diffeomorphic to the phase space
$T^*LG$ of the WZW model, which we obtained in the discussion
above (for example, the second homotopy groups differ).
The reason for this difference is that there is again
a Gribov-Singer ambiguity associated
with this gauge fixing; note that this is the one-dimensional analogue
of the four-dimensional Yang-Mills Gribov-Singer ambiguity [\gribov].
The $k$-symmetry just mentioned
can be further gauge fixed by choosing $A$ to be in the Cartan
subalgebra $\bf{h}$ of $\LieG$, and this
parameterisation was used together with a version of the $LR$ model to
calculate the Poisson brackets of this theory in Refs. [\chu, \us].

A requirement upon any choice of a space of solutions for a model
is that it should correspond to the Cauchy data for the model.
A solution of the equations of motion of the WZW model can be specified in a
neighbourhood of a Cauchy surface $S^1$ (say the Cauchy surface t=0) by the
Cauchy data $g(x,0)=f(x)$ and $(g^{-1}\partial_t g)(x,0)=w(x)$,
where $f$ and $w$ are
independent functions. If $u$ and $v$ have monodromy (as in Ref. [\chu]),
then the solution $g(x,t)= u(x^+) v(x^-)$ of
the WZW model has unconstrained Cauchy data (our solution \bthree\ similarily
has unconstrained Cauchy data). However, if
one requires that $u$ and $v$ are periodic (as in Ref.[\witten]),
then the Cauchy data carried by $f$ and $w$ is
constrained. This constraint is that the
holonomy of the connection $\half (f^{-1}\partial_x f- w)$ on the circle $S^1$
must be the identity group.

In conclusion, the parameterisation of the solutions of the WZW model given
in Section Three (Eqn. \bthree) is general in the sense that it is invariant
under a larger symmetry than other parameterisations considered
in the literature, and the latter can be thought of as locally-valid
gauge-fixed versions of it.
In our parameterisation, the covariant canonical phase space of the WZW model
is the same as the Hamiltonian phase space of the theory, and the calculation
of the Poisson brackets is straightforward.

\noindent{\bf Acknowledgements:} This work was supported by the SERC.

\refout

\bye